\documentclass{article}
\usepackage{arxiv}
\usepackage[utf8]{inputenc} % allow utf-8 input
\usepackage[T1]{fontenc}    % use 8-bit T1 fonts
\usepackage{hyperref}       % hyperlinks
\usepackage{url}            % simple URL typesetting
\usepackage{booktabs}       % professional-quality tables
\usepackage{amsfonts}       % blackboard math symbols
\usepackage{nicefrac}       % compact symbols for 1/2, etc.
\usepackage{microtype}      % microtypography
\usepackage{graphicx}
\usepackage{natbib}
\usepackage{doi}
\usepackage{float}
\usepackage{placeins}
\usepackage[font=small, skip=2pt]{caption}
\usepackage{graphicx}
\usepackage{caption}
\usepackage{subcaption}
\usepackage{booktabs}
\usepackage{array}
\usepackage{tikz}
\usepackage{titlesec}

\title{THE EMERGING AI \emph{'RÉVOLUTION TRANQUILLE'} IN AMERICA}

\date{May 19, 2025}	% Here you can change the date presented in the paper title

\author{ \href{https://orcid.org/0000-0003-0704-7787}{\includegraphics[scale=0.06]{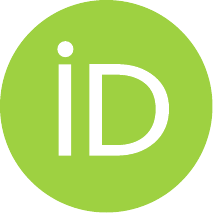}\hspace{1mm}Omar R. Malik}\\
	School of Management\\
	Kettering University\\
	Flint, MI 48504 \\
	\texttt{omalik@kettering.edu} \\
}

\renewcommand{\headeright}{Working Paper}
\renewcommand{\undertitle}{Working Paper}
\renewcommand{\shorttitle}{AI Adoption Trends in the US}

\hypersetup{
pdftitle={The emerging AI 'revoltion tranquille' in America},
pdfauthor={Omar R. Malik},
pdfkeywords={AI, Adoption, United States},
}

\begin{document}
\maketitle

\begin{abstract}

Using data from the U.S. Census Bureau’s Business Trends and Outlook Survey (BTOS), I examine the adoption of AI among US firms at national, state, industry, and firm size levels. I find that adoption remains overall low (only around 7\% of firms currently use AI), but is on a steady upward trajectory with a rising share of firms planning to implement AI. Adoption rates vary significantly across regions and sectors: some states are emerging as early adopters, while others lag, and knowledge-intensive industries (such as information technology and professional services) along with larger firms show higher openness to AI adoption compared to sectors like construction or small businesses. In general, these trends indicate that a “\textit{quiet revolution}” in AI adoption is underway; a gradual but expanding diffusion of AI across the economy with important implications for future productivity and policy.

\end{abstract}

\keywords{AI \and Adoption \and United States}

\titlespacing*{\section}{0pt}{*0.8}{*0.5}
\section{Introduction}

Artificial intelligence (AI) has emerged as a general-purpose technology (GPT) with the potential to significantly enhance reasoning and decision-making processes. Although AI is often colloquially associated with large language models (LLMs) and deep neural networks that power them, the broader domain of AI encompasses a wide range of computational tools, including machine learning, statistical analysis, and visualization methods that support the automation of analytical and cognitive tasks. The large-scale adoption of such systems has significant implications for productivity and economic growth at the national, state, and sectoral levels in the United States.

Despite rapid progress, AI systems remain imperfect. They may make incorrect inferences, hallucinate facts, or reproduce bias embedded in their training data. However, successive iterations have shown considerable improvements. Where earlier systems were criticized for lacking capacities in causal and counterfactual reasoning~\cite{10.1145/3722138}, newer models increasingly exhibit emergent reasoning abilities, including the ability to articulate chains of thought and engage in counterfactual logic. As these systems evolve, they are likely to become reliable complements or, in some cases, substitutes for human reasoning and decision making~\cite{NAP27644}.

This evolution is particularly consequential for knowledge workers, whose roles have traditionally involved diagnosing problems, inferring conclusions, and deciding on interventions~\cite{Mari}. These professionals, teachers, securities traders, analysts, and lawyers are central to modern service economies, and their activities account for almost half of the output generation in advanced economies~\cite{eig2015newmap}. As AI systems begin to perform tasks associated with these roles, the structure and value of knowledge work are likely to change significantly~\cite{Brynjolfsson, NAP27644}.

Given these stakes, understanding the pace and distribution of AI adoption is critical, not only from a technological standpoint, but also from its socio-economic and policy implications. To that end, the U.S. Census Bureau has begun collecting representative data on AI adoption across firms. These data provide an unprecedented opportunity to examine key questions: What are the national trends in the adoption of AI? How does adoption vary by region, industry, and firm size? What sectors and metropolitan areas are leading or lagging in this technological transition?

In this article, I use these data to answer these questions. My analysis highlights the rate, scope, and geographic distribution of AI adoption in the US economy, providing information on how this transformative technology is beginning to shape knowledge work in what remains the world’s most advanced service-based economy. My analysis shows that overall AI adoption is low but is increasing. There are emerging differences within regions in the adoption of AI and intentions to use AI. There are greater differences between industrial sectors in terms of AI adoption. Larger firms in knowledge-intensive sectors are adopting AI faster. Finally, cities in the West are leading the way in the adoption of AI. 

\titlespacing*{\section}{0pt}{*0.8}{*0.5}
\section{Data}

Since September 2023, the US Census Bureau has included questions related to artificial intelligence (AI) adoption in its biweekly \textit{Business Trends and Outlook Survey} (BTOS). Two recurring questions track AI adoption trends: whether a firm has used AI to develop goods or services in the past two weeks and whether it intends to use AI for those purposes within the next six months.

 BTOS surveys approximately 200,000 businesses every two weeks, with an average response rate of 16\% for AI-related questions. Thus, this analysis is based on aggregated responses from approximately 164,500 companies. The survey employs stratified random sampling, in which businesses of all sizes - small, medium, and large - are selected with equal probability. This ensures that the data set is broadly representative across the firm size distribution.

To compute aggregate adoption metrics, the Census Bureau applies survey weights that adjust for non-response and sampling design. The weighted responses are used to estimate the percentage of firms responding 'yes' or 'no' to the AI questions at the national, regional, sectoral, and firm size levels. Additional details on the survey design, sampling frame, and methodology are available from the US Census Bureau.\footnote{\url{https://www.census.gov/data/experimental-data-products/business-trends-and-outlook-survey.html}}

\section{National Trends in AI Adoption}

To assess trends in organizational use of artificial intelligence (AI), the US Census Bureau includes two recurring questions in its Business Trends and Outlook Survey (BTOS). Firms are asked whether they have used AI\footnote{AI is defined as ``computer systems and software that are able to perform tasks normally requiring human intelligence, such as decision-making, visual perception, speech recognition, and language processing.'' Examples include machine learning, natural language processing, virtual agents, and voice recognition systems.} in the production of goods and services during the last two weeks and if they intend to do so in the next six months.

\begin{figure}[H]
    \centering
    \caption{National trends in AI use and intent}
    \label{fig:1}
    \begin{subfigure}[t]{0.48\textwidth}
        \centering
        \includegraphics[width=\linewidth]{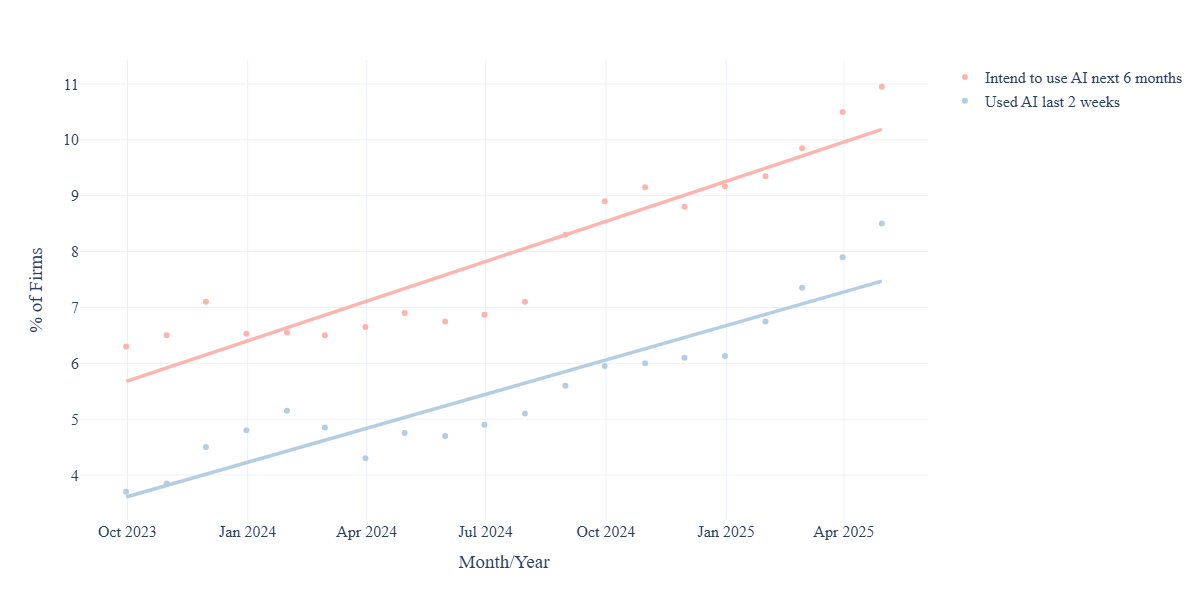}
        \caption{Used and intend to use AI: Yes}
        \label{fig:1a}
    \end{subfigure}
    \hfill
    \begin{subfigure}[t]{0.48\textwidth}
        \centering
        \includegraphics[width=\linewidth]{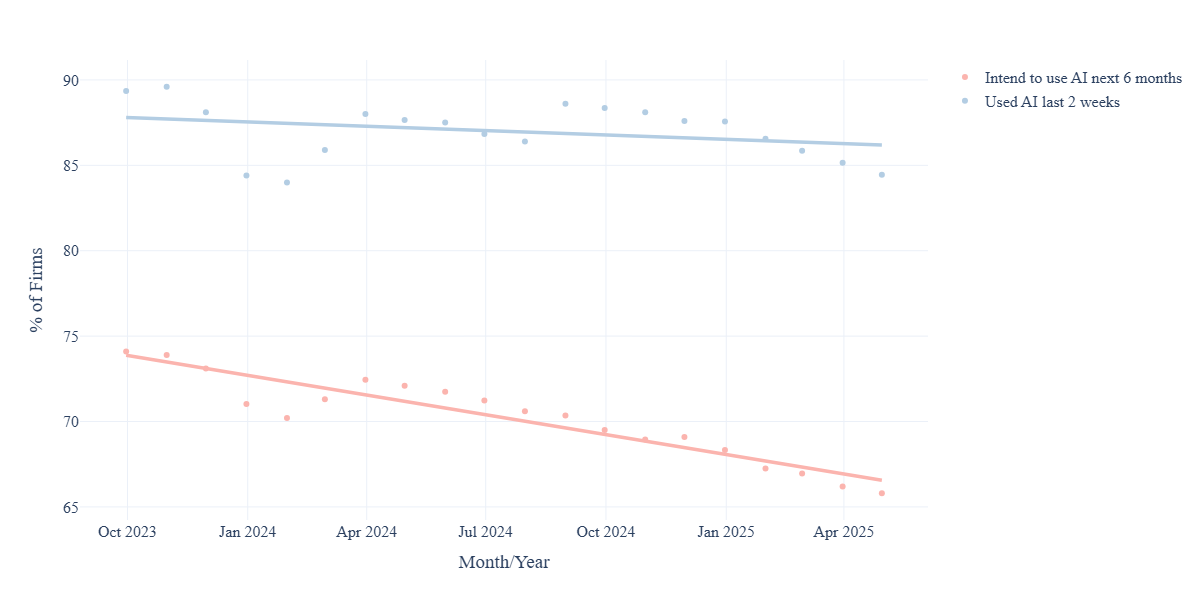}
        \caption{Used and intend to use AI: No}
        \label{fig:1b}
    \end{subfigure}
\end{figure}

The share of companies that responded positively to these questions has increased steadily over the survey period. As shown in Figure~\ref{fig:1a}, the reported use of AI increased from approximately 4\% to just above 7\%, while the intention to use AI increased from 6\% to 11\%. A linear trend explains 86\% the variance in these responses. Despite this growth, the overall prevalence of AI adoption remains modest, indicating that most US firms are in the early stages of exploring AI integration.

In contrast, a high proportion of companies continue to report no engagement with AI. As seen in Figure~\ref{fig:1b}, the share of companies that do not use AI remains close to 86\%, while those that do not intend to adopt AI in six months have decreased from 74\% to 66\%. These figures suggest that many firms have not yet identified compelling business cases for the implementation of AI. Broader adoption is likely to depend on clearer demonstrations of the value of AI in improving decision making and operational efficiency \cite{agrawal2022power}.

These findings contrast sharply with those reported by McKinsey \cite{singla2025stateofai}, indicating that 78\% of the largest companies globally had adopted AI in at least one function by 2024. The divergence is attributable in part to the sample composition: McKinsey's data are global and skewed toward large enterprises.\footnote{According to McKinsey, 42\% of survey respondents are from firms with annual revenues exceeding \$500 million.} This comparison underscores a bifurcation in AI adoption, with large firms moving more aggressively while smaller U.S. businesses remain in a preliminary exploratory phase.

\begin{figure}[H]
    \centering
    \caption{Used AI and intend to use AI: Don't know}
    \label{fig:3}
    \includegraphics[scale=0.35]{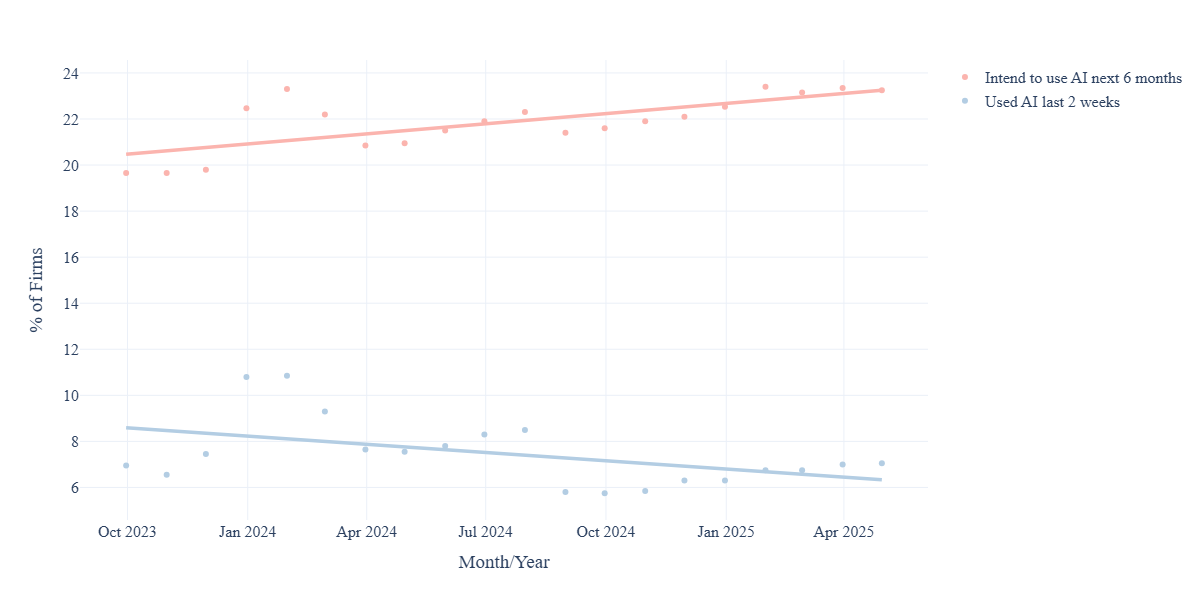}
\end{figure}

A notable proportion of firms indicate uncertainty about their use or planned use of AI. Approximately 7\% consistently report not knowing whether they used AI in the previous two weeks, and around 20\% are uncertain about their adoption plans in six months (Figure~\ref{fig:3}). This uncertainty may reflect a gap between informal, employee-level use of AI tools and formal, organization-wide integration into business processes.

In sum, while survey data reveal a clear upward trend in AI awareness and intent, they also highlight the persistent ambiguity and caution surrounding enterprise-level adoption. These dynamics suggest that AI integration in the US business landscape remains uneven and in early development.

\section{State Trends in AI Adoption}

The BTOS data also enable a detailed view of AI adoption in US states, providing insight into geographic variation in usage and intent. This section addresses several key questions: Do state-level patterns differ meaningfully from national trends? Are some states, particularly within regions, emerging as leaders or laggards in AI adoption? Are there signs of regional divergence in the adoption trajectory?

\begin{figure}[H]
    \centering
    \caption{Used AI in the last two Weeks}
    \begin{subfigure}[t]{0.48\textwidth}
        \centering
        \includegraphics[width=\linewidth]{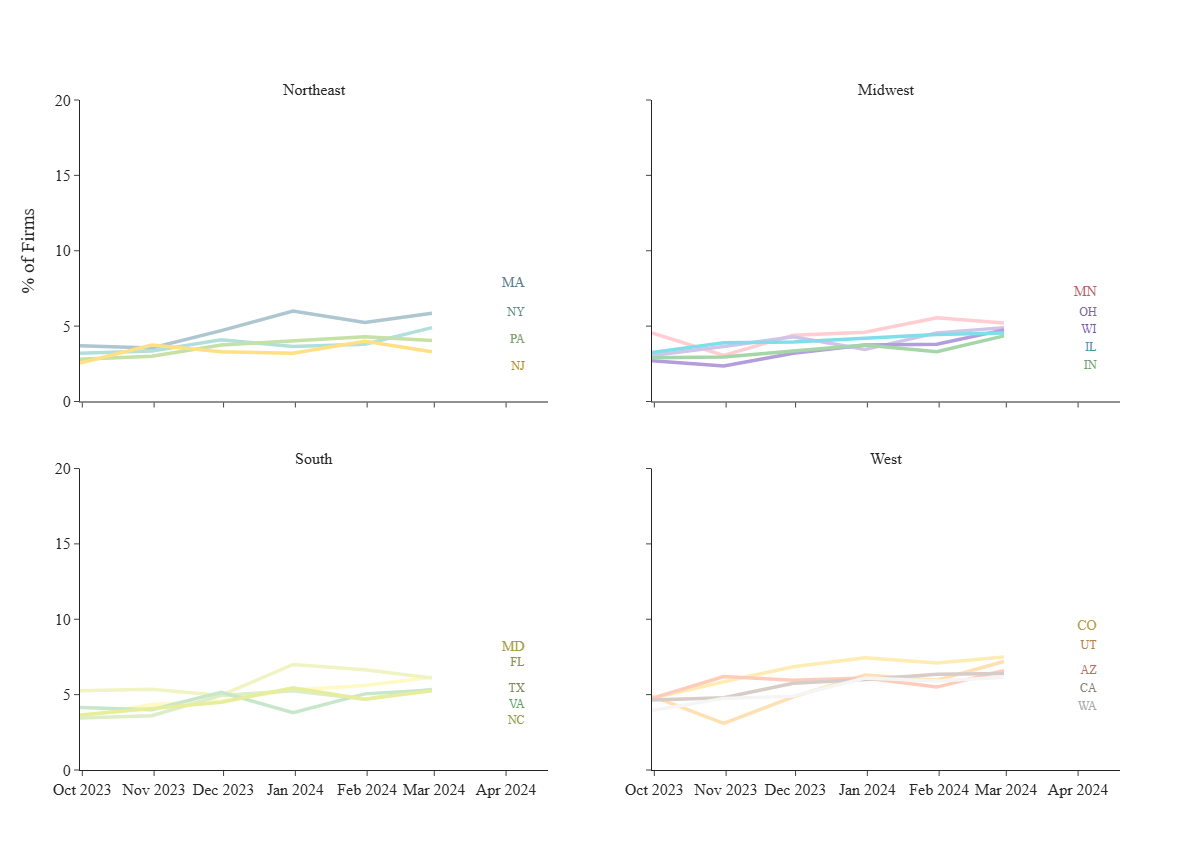}
        \caption{Used AI last two weeks: Yes}
        \label{fig:4}
    \end{subfigure}
    \hfill
    \begin{subfigure}[t]{0.48\textwidth}
        \centering
        \includegraphics[width=\linewidth]{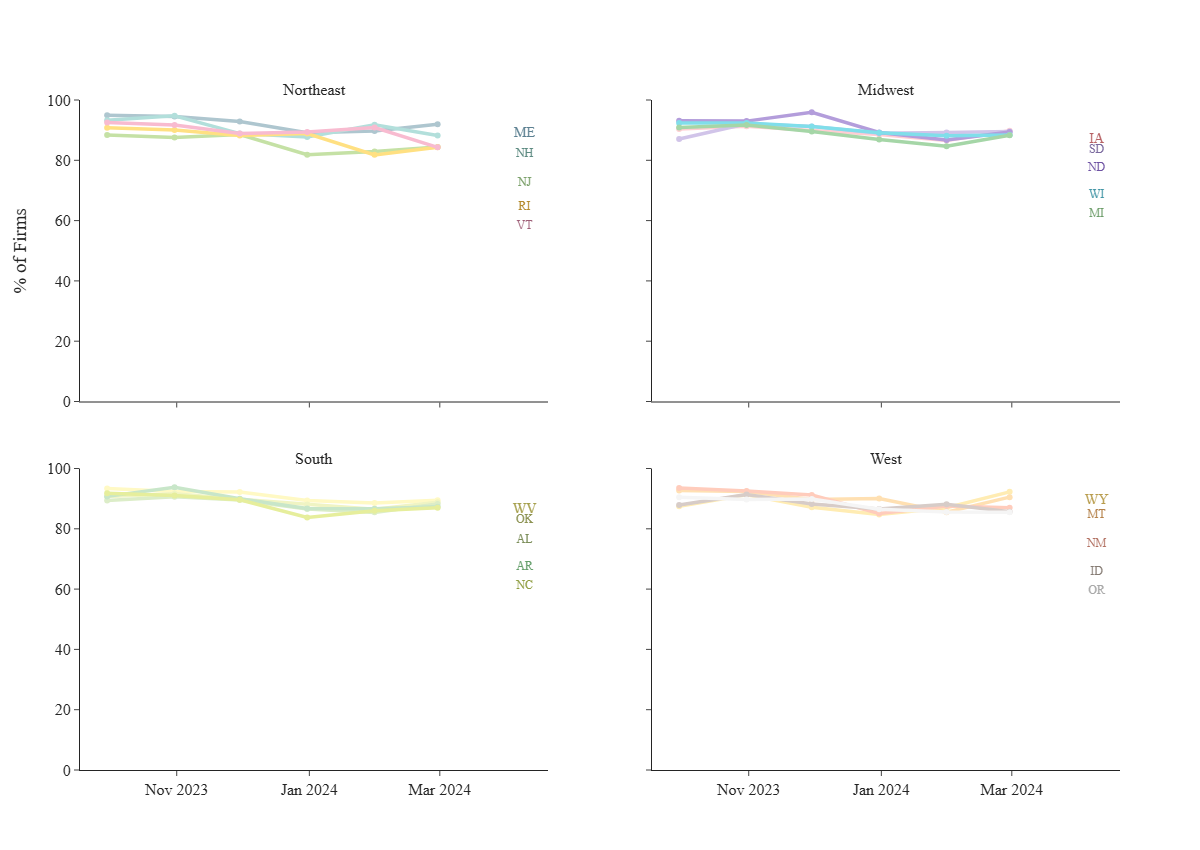}
        \caption{Used AI last two weeks: No}
        \label{fig:5}
    \end{subfigure}
\end{figure}

Across the four US Census regions, the share of firms that reported using AI in the past two weeks remains relatively low. As expected, the western region, home to many technology-focused firms and research institutions, leads in the current use of AI. Figure~\ref{fig:4} highlights the top five states in each region according to reported AI use. Although some states stand out, overall rates remain modest and the pace of growth is gradual. This suggests that most firms still do not view AI as a short-term productivity-enhancing technology, a trend that aligns with national patterns and the early-stage nature of AI adoption in a more general way \cite{agrawal2022power}.
 
\FloatBarrier

Figure \ref{fig:5} shows the top five states in each region, with firms saying no to using AI in the last two weeks. These percentages are high and consistent across the US regions. Large percentages of US firms, across all regions, remain skeptical of adopting AI. The rate of saying no to using AI is declining, but at a low rate.

\begin{table}[H]
\centering
\caption{Top 5 rankings of states by using AI last two Weeks}
\label{tab:used_ai_last_2weeks}
\begin{tabular}{l p{5cm} p{5cm}}
\toprule
Region & Used AI: Yes & Used AI: No \\
\midrule
Northeast &
\begin{tabular}[t]{@{}l@{}}
1. Massachusetts \\ 
2. New York \\ 
3. Pennsylvania \\ 
4. New Jersey
\end{tabular}
&
\begin{tabular}[t]{@{}l@{}}
1. Maine \\ 
2. New Hampshire \\ 
3. Vermont \\ 
4. Pennsylvania \\ 
5. Rhode Island
\end{tabular}
\\
\addlinespace
Midwest &
\begin{tabular}[t]{@{}l@{}}
1. Minnesota \\ 
2. Illinois \\ 
3. Ohio \\ 
4. Indiana \\ 
5. Wisconsin
\end{tabular}
&
\begin{tabular}[t]{@{}l@{}}
1. North Dakota \\ 
2. Wisconsin \\ 
3. South Dakota \\ 
4. Indiana \\ 
5. Iowa
\end{tabular}
\\
\addlinespace
South &
\begin{tabular}[t]{@{}l@{}}
1. Florida \\ 
2. Maryland \\ 
3. North Carolina \\ 
4. Virginia \\ 
5. Texas
\end{tabular}
&
\begin{tabular}[t]{@{}l@{}}
1. West Virginia \\ 
2. Oklahoma \\ 
3. Arkansas \\ 
4. Mississippi \\ 
5. Louisiana
\end{tabular}
\\
\addlinespace
West &
\begin{tabular}[t]{@{}l@{}}
1. Colorado \\ 
2. Arizona \\ 
3. California \\ 
4. Utah \\ 
5. Oregon
\end{tabular}
&
\begin{tabular}[t]{@{}l@{}}
1. Montana \\ 
2. Alaska \\ 
3. New Mexico \\ 
4. Wyoming \\ 
5. Idaho
\end{tabular}
\\
\bottomrule
\end{tabular}
\end{table}

Table \ref{tab:used_ai_last_2weeks} shows the top five ranked states based on the mean percentage from September 2023 to April 2025. These rankings hint at the divergence of AI within regions. Some states in each region are leading in AI use, while others seem to be lagging. However, there is some overlap between the rankings; for example, Pennsylvania and Indiana have firms that say Yes and No. At this point, percentage differences between states or regions are small and not meaningful.  

\begin{figure}[H]
    \centering
    \caption{Intend to use AI in the next six months}
    \begin{subfigure}[t]{0.48\textwidth}
        \centering
        \includegraphics[width=\linewidth]{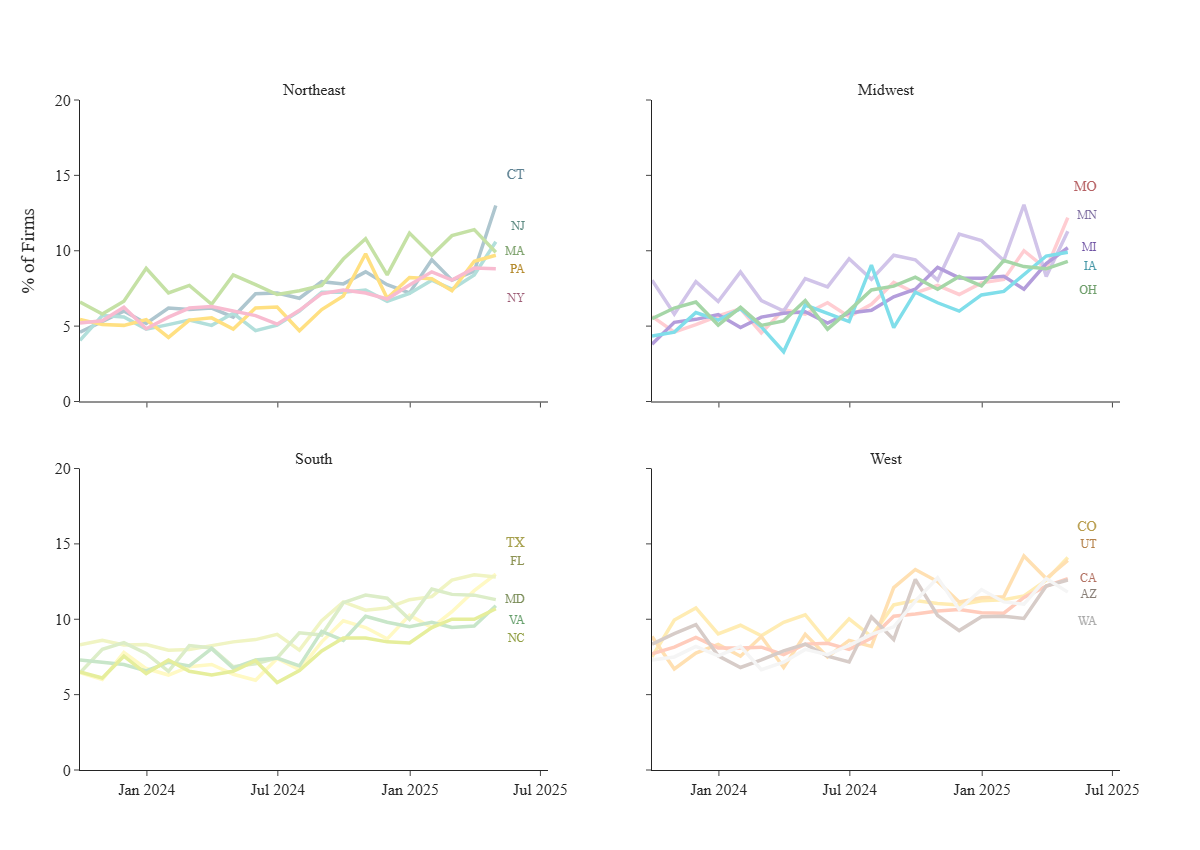}
        \caption{Intent to use AI: Yes}
        \label{fig:4a}
    \end{subfigure}
    \hfill
    \begin{subfigure}[t]{0.48\textwidth}
        \centering
        \includegraphics[width=\linewidth]{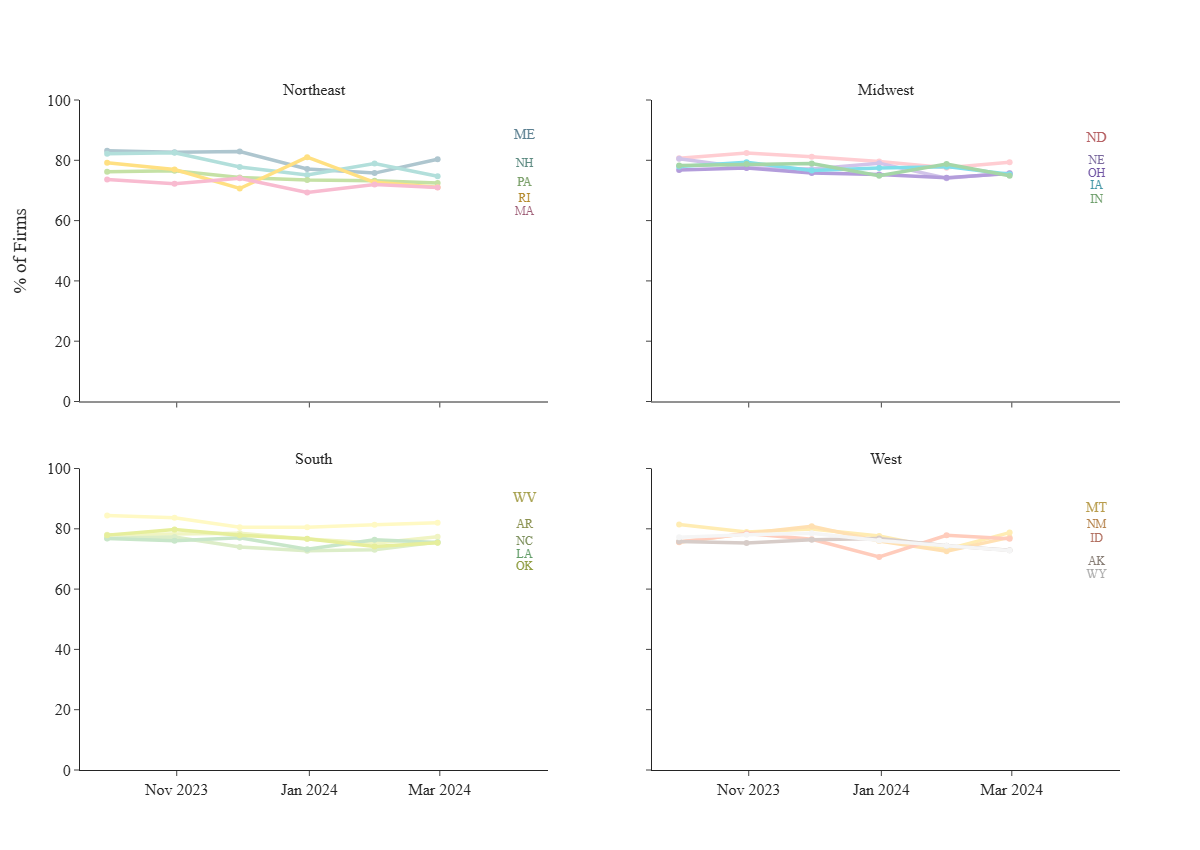}
        \caption{Intent to use AI: No}
        \label{fig:4b}
    \end{subfigure}
\end{figure}

A slightly more optimistic set of trends is visible in the responses to the question. Do you intend to use AI over the next 6 months to develop goods and services? More firms intend to use AI than firms that use AI, although at a fairly low rate, the highest percentages between states are between 7\% and 9\% (see Figure \ref{fig:4a}). However, the intention to not use AI remains also quite high. Between 70\% and 80\% firms in the states shown in Figure \ref{fig:4b} are saying 'No' to AI.  

\renewcommand{\tablename}{\textbf{Table}} % Makes 'Table 1' bold
\begin{table}[H]
\centering
\caption{Top 5 rankings of states by intent to use AI}
\label{tab:intent_ai_rankings}
\begin{tabular}{l p{5cm} p{5cm}}
\toprule
Region & Intent to Use AI: Yes & Intent to Use AI: No \\
\midrule
Northeast &
\begin{tabular}[t]{@{}l@{}}
1. Massachusetts \\ 
2. Connecticut \\ 
3. New York \\ 
4. New Jersey \\ 
5. Pennsylvania
\end{tabular}
&
\begin{tabular}[t]{@{}l@{}}
1. Maine \\ 
2. New Hampshire \\ 
3. Vermont \\ 
4. Rhode Island \\ 
5. Pennsylvania
\end{tabular}
\\
\addlinespace
Midwest &
\begin{tabular}[t]{@{}l@{}}
1. Minnesota \\ 
2. Illinois \\ 
3. Ohio \\ 
4. Missouri \\ 
5. Indiana
\end{tabular}
&
\begin{tabular}[t]{@{}l@{}}
1. North Dakota \\ 
2. South Dakota \\ 
3. Iowa \\ 
4. Indiana \\ 
5. Wisconsin
\end{tabular}
\\
\addlinespace
South &
\begin{tabular}[t]{@{}l@{}}
1. Florida \\ 
2. Maryland \\ 
3. Virginia \\ 
4. Texas \\ 
5. North Carolina
\end{tabular}
&
\begin{tabular}[t]{@{}l@{}}
1. West Virginia \\ 
2. Mississippi \\ 
3. Arkansas \\ 
4. Oklahoma \\ 
5. Alabama
\end{tabular}
\\
\addlinespace
West &
\begin{tabular}[t]{@{}l@{}}
1. Colorado \\ 
2. Nevada \\ 
3. Arizona \\ 
4. California \\ 
5. Utah
\end{tabular}
&
\begin{tabular}[t]{@{}l@{}}
1. Montana \\ 
2. New Mexico \\ 
3. Wyoming \\ 
4. Idaho \\ 
5. Alaska
\end{tabular}
\\
\bottomrule
\end{tabular}
\end{table}

Table~\ref{tab:intent_ai_rankings} presents the top five states in each region according to the stated intent of the firms to use AI. In most regions, there is little overlap between states ranking high in intent and those ranking high in non-intent. In particular, Pennsylvania appears on the 'Yes' and 'No' lists in the Northeast, and Indiana appears on both lists in the Midwest. In contrast, the South and West show no overlap, suggesting a clearer internal divergence. These patterns suggest that even within regions, states are on different trajectories with respect to AI adoption. Although current differences remain modest, a growing divergence could lead to significant disparities in economic opportunity and technological competitiveness. To explore this dynamic further, Figure~\ref{fig:ai_use_grid} offers a classification of states based on the combined current use and future intent.

\begin{figure}[H]
\centering
\caption{Firms in States: AI use and intent to use}
\label{fig:ai_use_grid}
\begin{tikzpicture}[font=\scriptsize]
  % Define cell size
  \def\cellwidth{5cm}
  \def\cellheight{2.5cm}

  % Draw the grid
  \draw (0,0) rectangle ({2*\cellwidth},{1.8*\cellheight});
  \draw (\cellwidth,0) -- (\cellwidth,{1.8*\cellheight});
  \draw (0,\cellheight) -- ({2*\cellwidth},\cellheight);

  % X-axis labels
  \node at ({\cellwidth/2},{2*\cellheight + 0.3}) {No};
  \node at ({1.5*\cellwidth},{2*\cellheight + 0.3}) {Yes};
  \node at (\cellwidth,{2*\cellheight + 0.8}) {\textbf{Used AI (Last 2 Weeks)}};

  % Y-axis labels
  \node[rotate=90] at (-0.8,{1.0*\cellheight}) {\textbf{Intent to Use AI (Next 6 Months)}};
  \node at (-0.3,{\cellheight/2}) {No};
  \node at (-0.3,{1.5*\cellheight}) {Yes};

  % Cell A: Intent Yes, Used No
  \node[align=center, text width=4.5cm, anchor=center] at ({\cellwidth/2},{1.5*\cellheight}) {
    IA, PA
  };

  % Cell B: Intent Yes, Used Yes
  \node[align=center, text width=4.5cm, anchor=center] at ({1.5*\cellwidth},{1.5*\cellheight}) {
    AZ, CA, CO, FL, IL, IN, MA, MD, MN, NC, NJ, NY, OH, PA, TX, UT, VA
  };  

  % Cell C: Intent No, Used No
  \node[align=center, text width=4.5cm, anchor=center] at ({\cellwidth/2},{\cellheight/2}) {
    AK, AR, IA, ID, IN, ME, MS, MT, ND, NH, NM, OK, PA, RI, SD, VT, WI, WV, WY
  };

  % Cell D: Intent No, Used Yes
  \node[align=center, text width=4.5cm, anchor=center] at ({1.5*\cellwidth},{\cellheight/2}) {
    IN, PA, WI
  };
\end{tikzpicture}
\end{figure}

Figure~\ref{fig:ai_use_grid} categorizes the states according to whether the firms report both using AI and intending to use it, only one of the two, or neither. The states in the upper right quadrant, where firms report both current use and future intent, can be considered \textit{early leaders} in AI adoption. The top left quadrant includes states where companies have not yet adopted AI but plan to do so in the near future; these may be seen as \textit{experimenters}. The bottom left quadrant contains states where firms neither use nor intend to use AI, indicating a group of potential \textit{laggards}. Finally, the bottom right quadrant comprises states where firms have adopted AI but do not plan to continue or expand use, suggesting a segment of \textit{skeptics}.

A particularly concerning group are the states that appear only in the bottom-left quadrant and are absent from all others: those where firms report no use and no intent to adopt AI. These represent the most persistent low-adoption zones. They include Alaska, Arkansas, Idaho, Iowa, Maine, Mississippi, Montana, New Hampshire, New Mexico, North Dakota, Rhode Island, South Dakota, Vermont, and Wyoming. Regionally, this includes four states in the Northeast, three in the Midwest, two in the South, and five in the West.

Unlike some nation-states that have adopted coordinated policies to build competitive advantage through AI, such as China, US states lack unified policy frameworks to promote firm-level AI adoption. Without targeted interventions, these geographic disparities can deepen, with long-term consequences for regional economic development and technological equity.

\section{Industrial Sectors and Firm Sizes}

\begin{figure}[H]
    \centering
    \caption{AI use In industrial sectors}
    \begin{subfigure}[t]{0.48\textwidth}
        \centering
        \includegraphics[width=\linewidth]{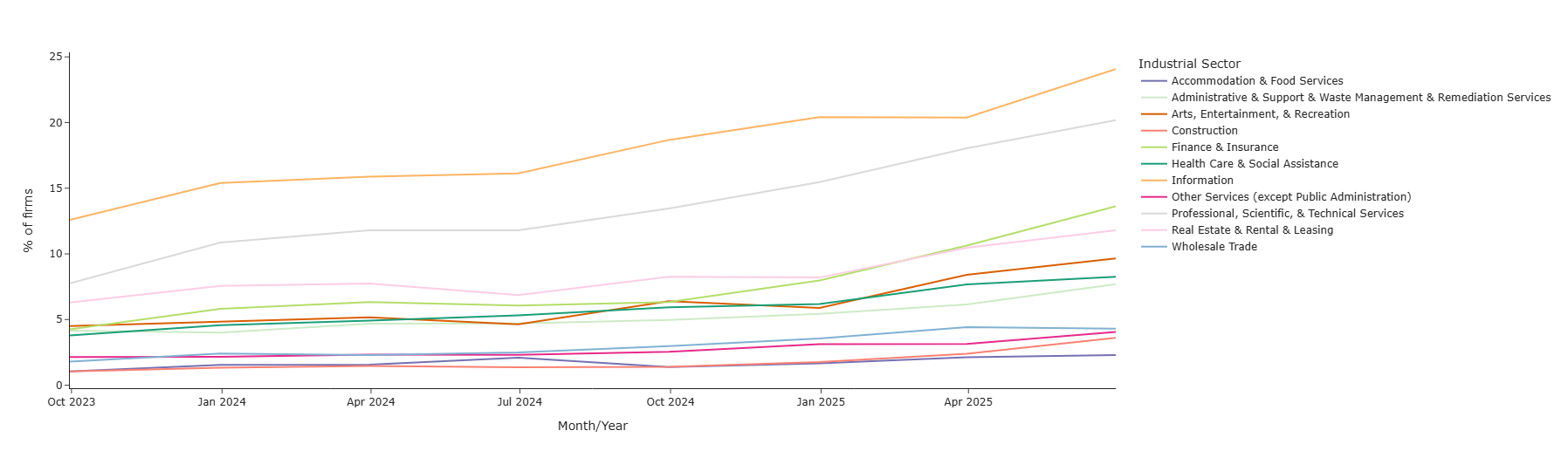}
        \caption{Used AI: Yes}
        \label{fig:6a}
    \end{subfigure}
    \hfill
    \begin{subfigure}[t]{0.48\textwidth}
        \centering
        \includegraphics[width=\linewidth]{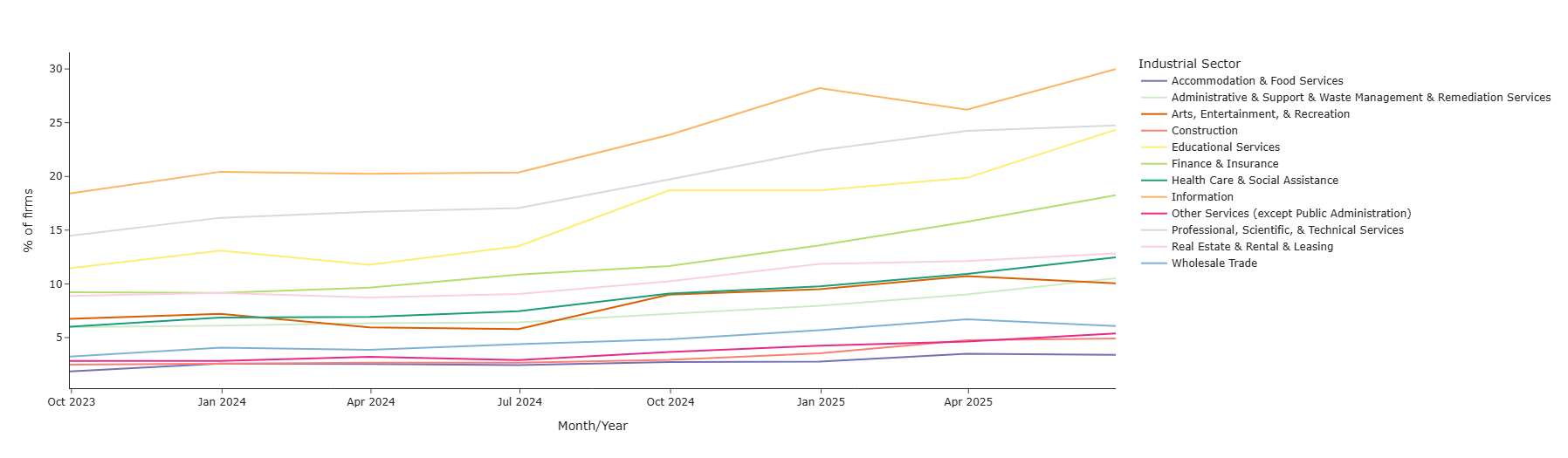}
        \caption{Intent to use AI: Yes}
        \label{fig:6b}
    \end{subfigure}
\end{figure}

The use of artificial intelligence is increasing in a wide range of US industrial sectors \cite{singla2025stateofai}. As can be expected, knowledge-intensive sectors such as Information Technology, Professional \& Technical services, Finance \& Insurance, and Real Estate, Rental \& Leasing are leading adopters of AI systems. Lagging sectors include Accommodation \& Food Services, Construction and Wholesale trade (see Figure \ref{fig:6a}). Although it is easy to imagine highly productive AI applications that could aid in key decision-making and reasoning processes in these industries. 

Figure \ref{fig:6b} shows very similar trends in industries that plan to use AI in the next six months. Across industries, over the course of the survey, the top five industries of firms that have used AI over the last two weeks\footnote{Mean percentage over the time frame of the survey.} are the following: Information technology, Professional, scientific and technical services, Real estate, rental, \& leasing, Finance and insurance, and arts, entertainment \& recreation.  The top industries for firms that intend to use AI are the same, except for adding educational services to the list. See Table \ref{tab:top5-industries-ai}

\begin{table}[H]
\centering
\caption{\textbf{Top 5 industries using/intending to use AI (\% of Firms)}}
\label{tab:top5-industries-ai}
\begin{tabular}{lcc}
\toprule
\textbf{Industry} & \textbf{Used AI \%} & \textbf{Intend to Use AI \%} \\
\midrule
Information & 15.59 & 20.15 \\
Professional Services & 11.01 & 16.22 \\
Educational Services & -- & 12.51 \\
Finance \& Insurance & 5.94 & 9.35 \\
Real Estate & 7.55 & 8.95 \\
\bottomrule
\end{tabular}
\end{table}

In contrast, the five lowest industries in terms of using AI and intending to use AI are shown in Table \ref{tab:smallest5-industries-ai}

\begin{table}[H]
\centering
\caption{\textbf{Bottom 5 industries using/intending to use AI (\% of Firms)}}
\label{tab:smallest5-industries-ai}
\begin{tabular}{lcc}
\toprule
\textbf{Industry} & \textbf{Used AI \%} & \textbf{Intend to Use AI \%} \\
\midrule
Construction & 1.39 & 2.55 \\
Accommodation/Food Services & 1.55 & 2.50 \\
Wholesale Trade & 2.26 & 3.85 \\
Other Services & 2.28 & 2.85 \\
Admin. Support, Waste Management, Remediation & 4.20 & 6.12 \\
\bottomrule
\end{tabular}
\end{table}

\begin{figure}[H]
    \centering
    \caption{Industrial sectors not using or intending to Use AI}
    \begin{subfigure}[t]{0.48\textwidth}
        \centering
        \includegraphics[width=\linewidth]{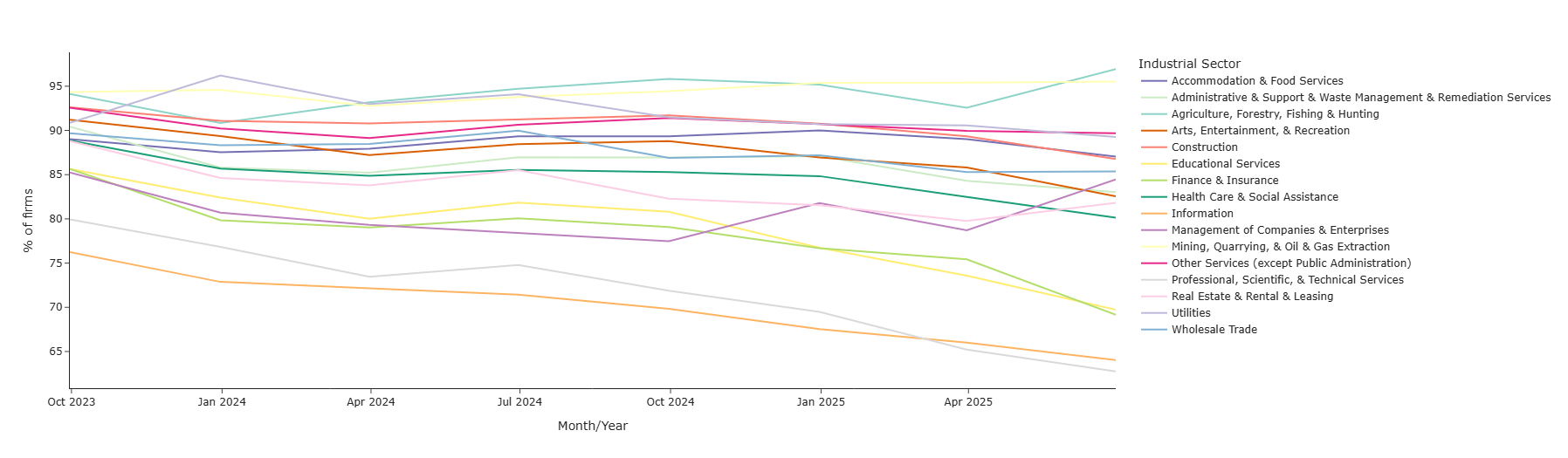}
        \caption{Used AI: No}
        \label{fig:7a}
    \end{subfigure}
    \hfill
    \begin{subfigure}[t]{0.48\textwidth}
        \centering
        \includegraphics[width=\linewidth]{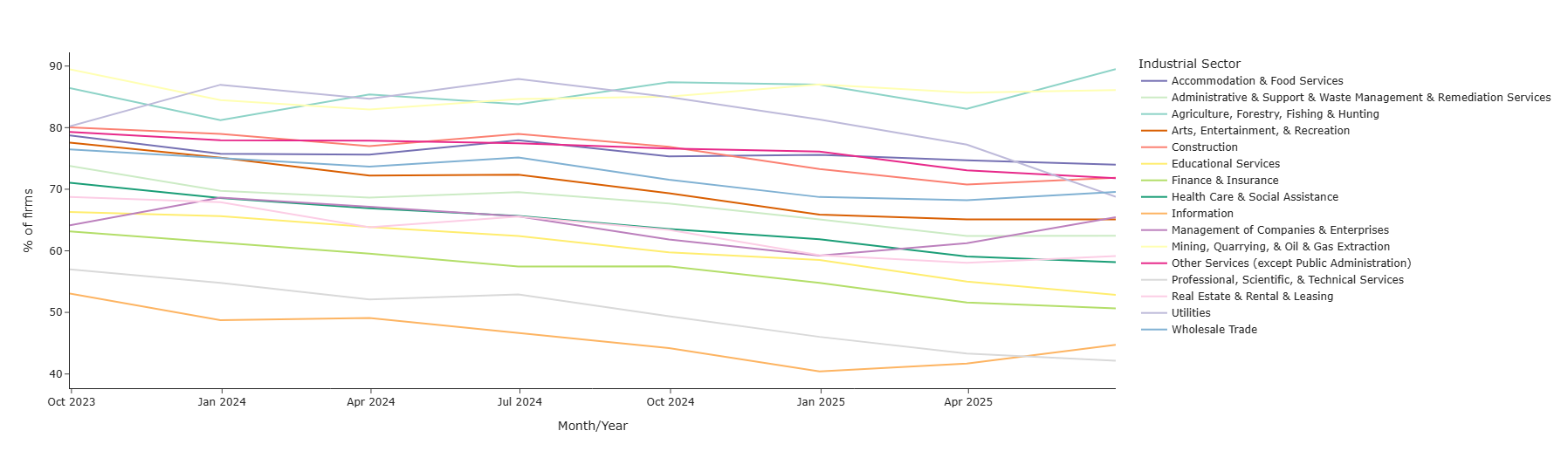}
        \caption{Intent to use AI: No}
        \label{fig:7b}
    \end{subfigure}
\end{figure}

Thus, certain industries can potentially be fast adopters and can gain from AI systems, while other industries are lagging behind. This will create a divergence in the gains from AI between industries over time. These trends are visible in Figure \ref{fig:7a}. Industries that are saying 'No' to using AI are high percentages of firms and stable over time, while lower lines like IT and Professional services, firms saying 'No' are declining over time. Very similar trends are visible in industries that do not intend to use AI in the next six months. Shown in Figure \ref{fig:7b}. 

\begin{figure}[H]
    \centering
    \caption{Firm size-industrial sectors using AI last two weeks: Yes}
    \label{fig:8}
    \includegraphics[scale=0.25]{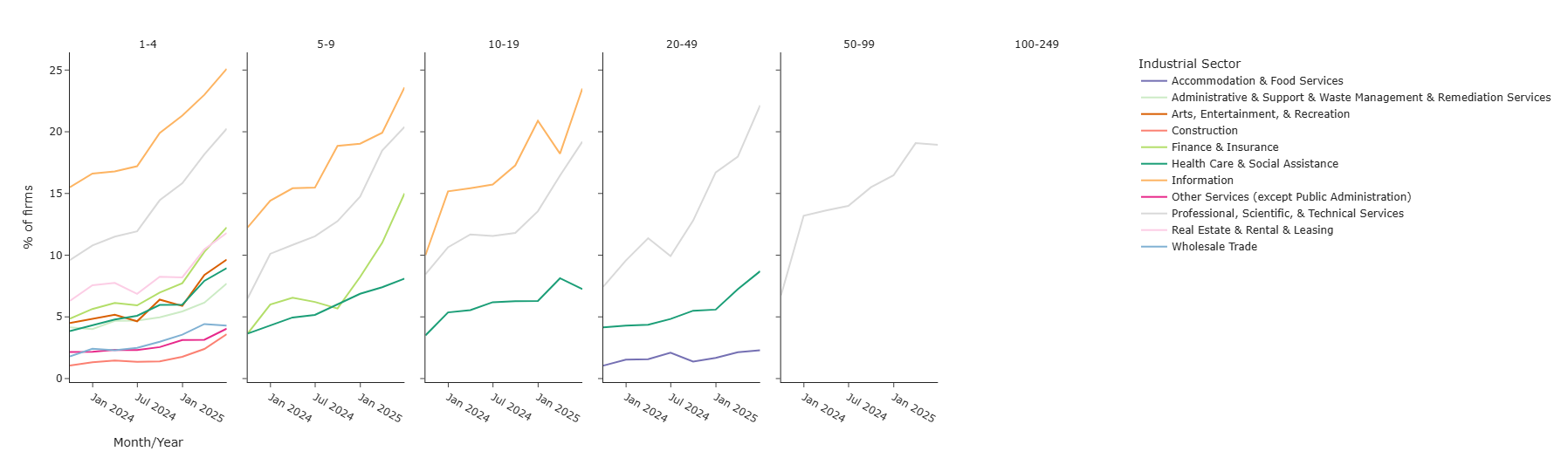}
\end{figure}

Next, I examine the interactions between firm size (measured as the number of employees), industrial sectors, and AI adoption. Figure \ref{fig:8} shows the same industrial sectors: IT and Professional Services, across firm sizes, are adopting AI the fastest. Another interesting trend is the increasing rate of AI adoption among small Finance \& Insurance firms. This is most noticeable in firms with 5-9 employees. Generally, across sizes and sectors, firms are using AI more over time, although data are sparser for firms larger than four employees.  

\begin{figure}[H]
    \centering
    \caption{Firm size-industrial sectors intending to use AI: Yes}
    \label{fig:9}
    \includegraphics[scale=0.25]{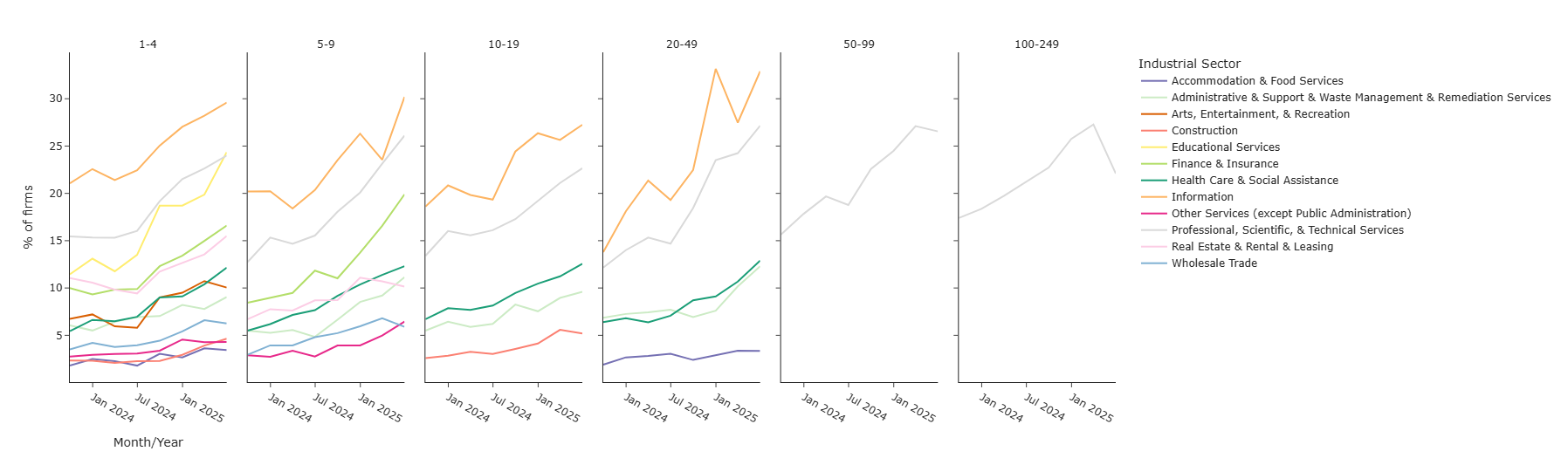}
\end{figure}

Figure \ref{fig:9} shows the intention to use AI over the next six months within industries and across firm sizes. These trends are very similar to firms that use AI. IT and professional services lead, while accommodation, other services, and arts lag behind. An interesting trend in the panel of 10-19 employees is the lagging intention to use AI in the construction sector. 

\begin{figure}[H]
    \centering
    \caption{Firm size-industrial sectors used AI last two weeks: No}
    \label{fig:10}
    \includegraphics[scale=0.25]{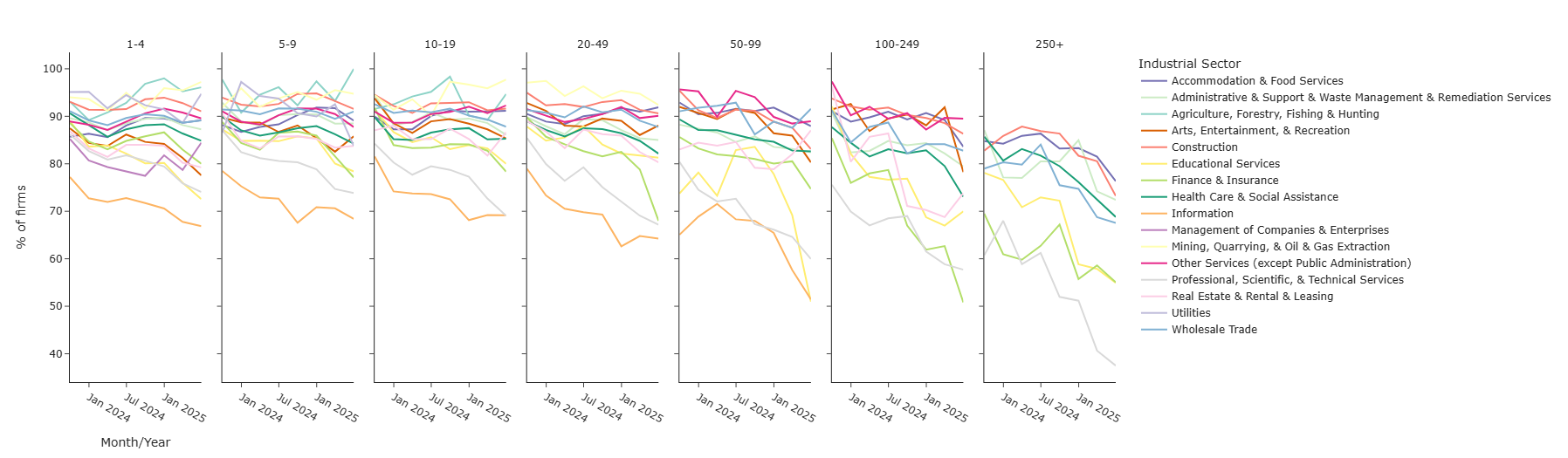}
\end{figure}

Figure \ref{fig:10} shows that between firm sizes and sectors, firms that did not use AI in the last two weeks are declining. These rates are falling faster for IT, Professional Services, Educational Services, and Real Estate firms. In addition, larger firms are displaying sharper declining rates than smaller firms. Thus, preliminary evidence suggests that larger firms in knowledge-intensive industrial sectors are opening up to adopt AI faster.  

At the same time, companies of various sizes in industrial sectors are declining in the percentage of saying no to AI adoption over the next six months. Once again, these trends are the strongest in IT and Professional Services. Interestingly, middle-sized educational services organizations (50-99 employees) also show an openness to using AI in service development. Larger firms in the health care, waste management, and construction sectors are also showing signs of decreasing intent to say no to AI.

\begin{figure}[H]
    \centering
    \caption{Firm size-industrial sectors intend to use AI next six months: No}
    \label{fig:11}
    \includegraphics[scale=0.25]{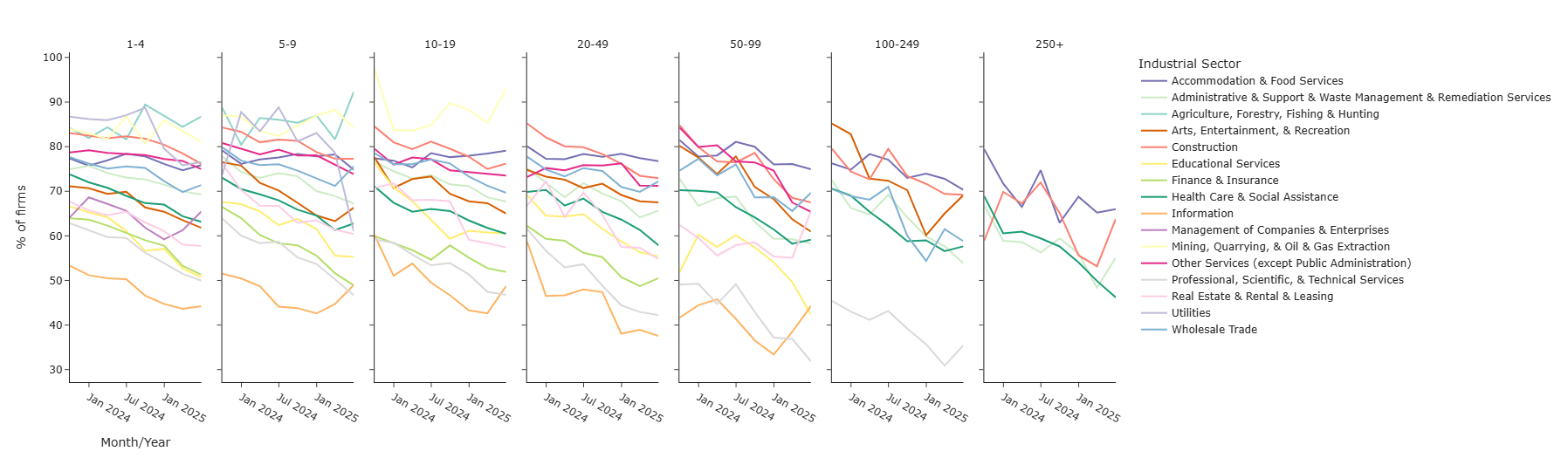}
\end{figure} 

Together, these trends show that IT and Professional Services, Finance \& Insurance and Rental \& Leasing are industrial sectors that are leading in AI adoption. Larger firms are more open to experimenting with AI, and across industries and firm sizes, AI adoption is increasing over time.

\section{Metro Areas}

A substantial share of US economic activity is concentrated in the country’s largest metropolitan statistical areas (MSAs), making them critical focal points for understanding technological adoption.  BTOS data offer valuable information on how AI adoption is evolving in these key urban centers.

\begin{figure}[H]
    \centering
    \caption{AI use in US MSAs}
    \begin{subfigure}[t]{0.48\textwidth}
        \centering
        \includegraphics[width=\linewidth]{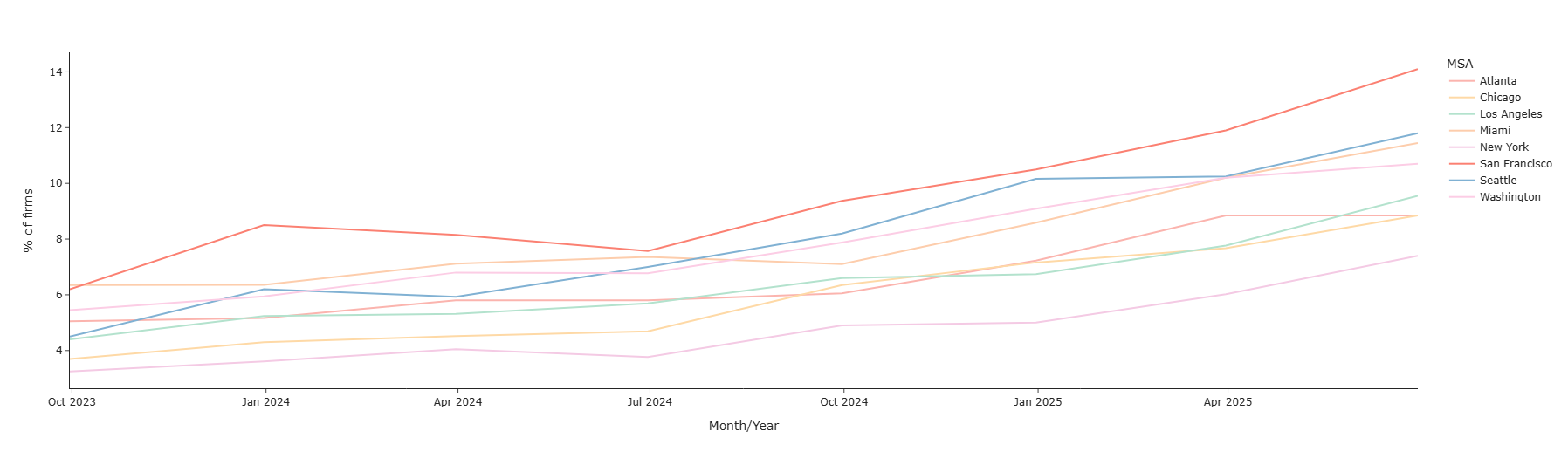}
        \caption{Used AI in the last two weeks: Yes}
        \label{fig:12a}
    \end{subfigure}
    \hfill
    \begin{subfigure}[t]{0.48\textwidth}
        \centering
        \includegraphics[width=\linewidth]{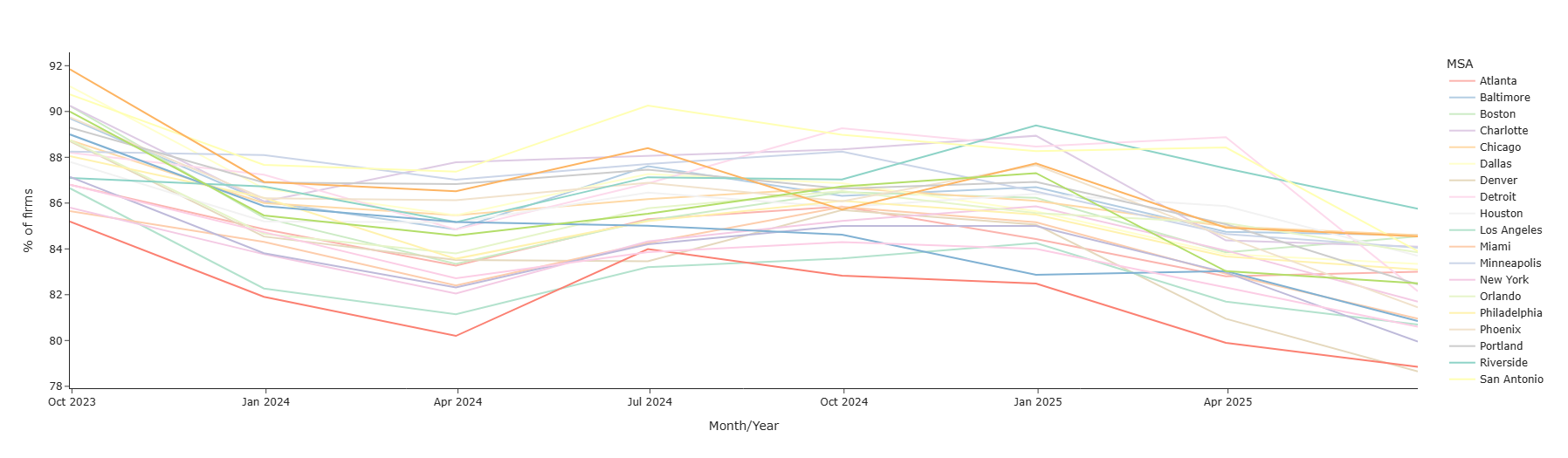}
        \caption{Used AI in the last two weeks: No}
        \label{fig:12b}
    \end{subfigure}
\end{figure}

As shown in Figures \ref{fig:12a} and \ref{fig:12b}, the adoption of AI is gradually increasing in most major MSAs, while the share of firms that report no use of AI is steadily declining. San Francisco, Seattle, and Miami stand out as the leading MSAs in terms of AI use, while cities such as Atlanta, Chicago, and New York exhibit comparatively slower growth. In particular, the most rapid declines in firms reporting non-use of AI are observed in Denver and San Francisco (Figure \ref{fig:12b}). Despite these positive trends, the overall share of companies that do not use AI remains high, between 78\% and 86\%, indicating that widespread adoption is still in its early stages.

\begin{figure}[H]
    \centering
    \caption{Intent to use AI in US MSAs}
    \begin{subfigure}[t]{0.48\textwidth}
        \centering
        \includegraphics[width=\linewidth]{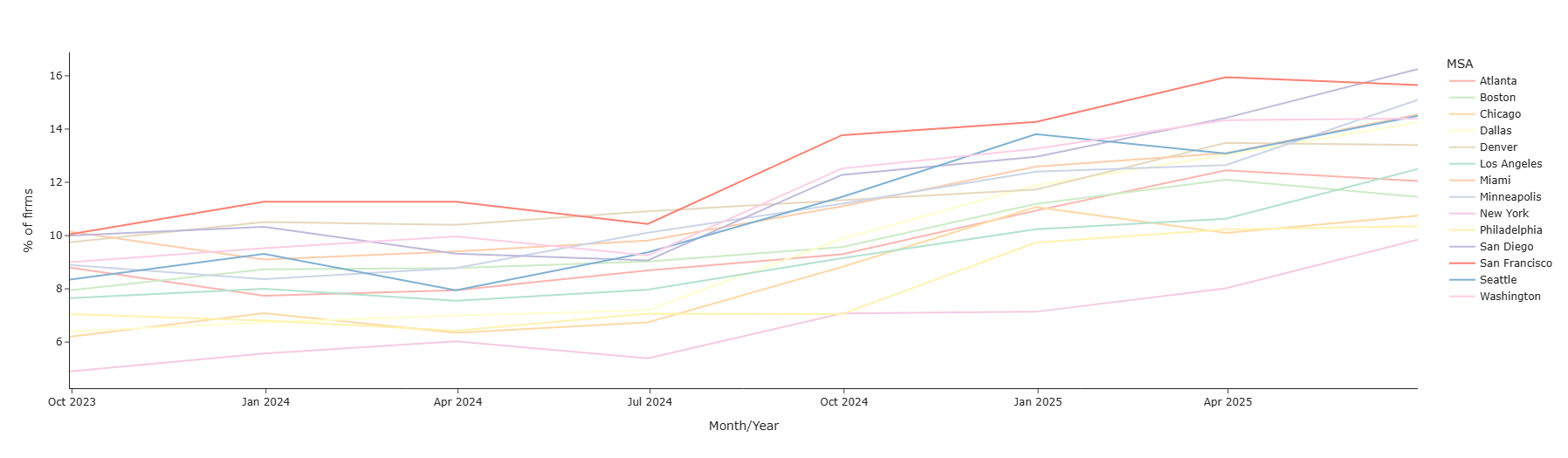}
        \caption{Intend to use AI next six months: Yes}
        \label{fig:13a}
    \end{subfigure}
    \hfill
    \begin{subfigure}[t]{0.48\textwidth}
        \centering
        \includegraphics[width=\linewidth]{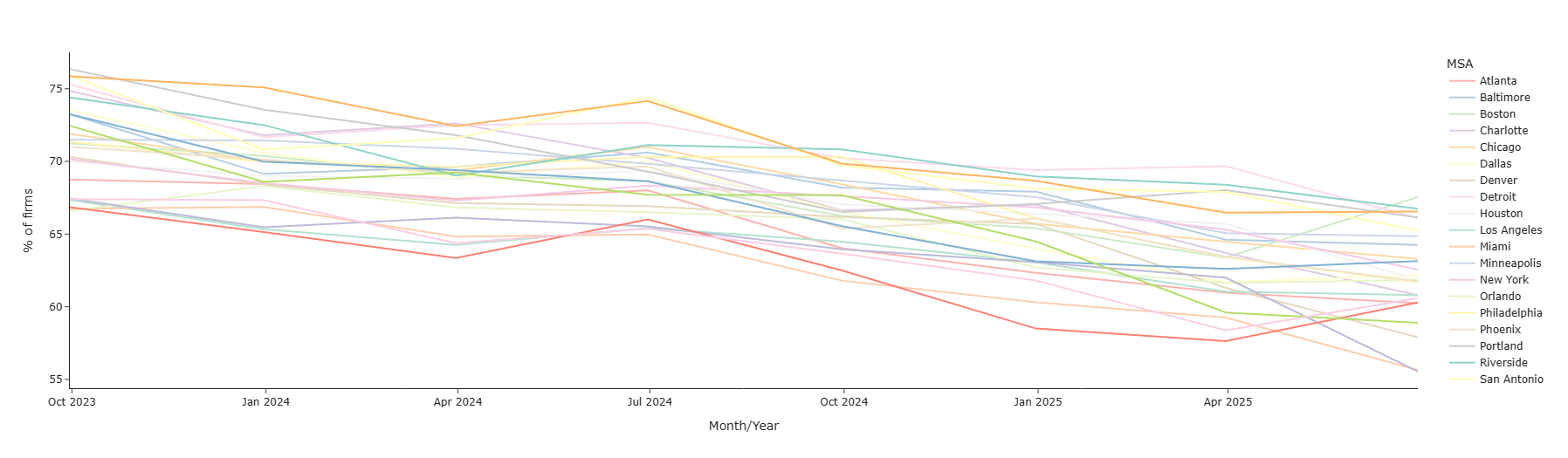}
        \caption{Intend to use AI next six months: No}
        \label{fig:13b}
    \end{subfigure}
\end{figure}

Trends in firms' intentions to adopt AI in US MSAs mirror the patterns observed in current usage. As shown in Figure \ref{fig:13a}, San Diego, San Francisco, Minneapolis, and Seattle emerge as the leading metropolitan areas where firms report plans to implement AI in the next six months. In contrast, although New York, Philadelphia, and Chicago exhibit some growth in intention, their adoption trajectories remain notably slower than those of the leading MSAs. These patterns highlight a growing divergence between firms on the East and West coasts. Figure \ref{fig:13b} illustrates the share of companies that do not intend to adopt AI within the next six months. Interestingly, these percentages are declining and at a faster rate than the declines observed for current non-use of AI. However, the intention to adopt AI remains limited, with 55\% to 65\% of the firms in the main MSAs still reporting no near-term plans for adoption.   

Collectively, these trends show that AI use and intentions of using AI are increasing in major US cities. However, a gap is opening between leading cities and some of the old-growth cities such as New York and Chicago. These gaps may portend a need for city policies that encourage responsible AI use by firms. 

\section{Discussion and Conclusion}

This study provides one of the first comprehensive looks at firm-level AI adoption across the United States using representative survey data. The key finding is that overall AI adoption remains low: only around 7\% of firms currently use AI – but usage is steadily growing. Moreover, there is an increasing share of firms (now around 11\%) reporting plans to implement AI in the near future. This suggests that the diffusion of AI is still in its early stages, yet clearly underway, constituting a \textit{'quiet revolution'} of gradual but significant technological uptake. The significance of this pattern lies in its parallel to past general-purpose technologies: Even transformative innovations often exhibit slow initial adoption before reaching a tipping point of higher diffusion \cite{agrawal2022power}. My findings indicate that AI is following this path, with important implications for productivity growth and competitive dynamics once adoption becomes more widespread.

Equally important are the substantial disparities in AI adoption between regions, industries, and firm sizes. Certain states and metropolitan areas have emerged as early leaders, notably tech-centric economies on the West Coast, while many others lag behind. Knowledge-intensive industries (e.g., information technology, professional and technical services) and larger firms show considerably higher adoption rates and intent to use AI, in contrast to sectors like construction, wholesale trade, or small businesses that remain largely on the sidelines. These disparities are significant because they foreshadow a widening gap in innovation-led growth. Firms and regions at the forefront of AI adoption can reap enormous productivity and efficiency gains, while lagging sectors and communities risk falling further behind. In sum, the evidence points to an emerging but uneven AI-driven transformation of the US economy – a \textit{'révolution tranquille'} in which change quietly gathers momentum rather than erupts overnight.

These findings have several implications for theories of technology diffusion and economic geography. First, the observed slow but steady uptake of AI aligns with the classic diffusion of innovation theory, which posits that new technologies often penetrate gradually – led by innovators and early adopters – before broader adoption occurs. The data confirm that AI adoption currently remains concentrated among a small minority of firms, consistent with the early phase of a diffusion S-curve. This underscores the argument \cite{Brynjolfsson} that general-purpose technologies require complementary investments and time to produce productivity rewards. My results reinforce the notion that AI, as a general-purpose technology, is following an incremental adoption path, necessitating updates to models of the productivity paradox and technology lifecycle to account for lag times in organizational uptake of AI. 

\section{Practical Implications}

From a practical point of view, this study offers several insights for business leaders and policy makers. For business professionals and managers, the low overall adoption rates mean that there is still a significant first-mover advantage to be captured. Early adopting firms, especially in lag industries, may gain competitive advantages and innovation benefits by integrating AI ahead of their peers. However, managers should also take into account the finding that many firms have yet to identify compelling use cases for AI. This implies a need for demonstrable ROI and successful pilot projects. Providers of AI solutions must bridge this gap by developing industry-specific applications and evidence of value. In practice, our results encourage firms to start with small-scale AI implementations to build internal capabilities and trust, given that even knowledge-intensive sectors are only gradually ramping up AI use. 

For policymakers, the uneven geography of AI adoption is a call to action. Unlike countries such as China that have coordinated national strategies, US states currently lack unified policy frameworks to promote AI diffusion. My findings suggest that without targeted support, such as incentives, training programs, or knowledge-sharing initiatives, less technologically advanced regions and industries could fall further behind in the AI economy. Policy interventions should focus on closing the adoption gap, for example, by supporting workforce development in AI skills, subsidizing adoption for small and midsize enterprises, and fostering regional innovation ecosystems. Moreover, the rising intent to adopt AI even in traditionally slower sectors (e.g. education, healthcare) indicates an opportunity for public-private partnerships to guide responsible and effective implementation. In summary, the practical implication is that achieving the benefits of AI throughout the economy will require proactive measures to ensure broader and more equitable adoption in industries and regions.

\section{Future Research Directions}

Future theoretical work could build on this evidence to refine predictions about when the AI adoption curve might inflect into more rapid growth. Second, the variation in adoption across regions and sectors informs theories of regional innovation systems and economic geography. The emergence of AI hotspots (for example, California, Washington, and select urban centers) versus low adoption zones (including parts of the Midwest, South, and rural areas) suggests that existing innovation centers are extending their lead in the AI era. This pattern aligns with cumulative advantage theories, where regions with strong human capital and tech ecosystems are better positioned to absorb new innovations. It also highlights the role of localized knowledge spillovers and policy support - or lack thereof - in shaping technology diffusion. The fact that some states appear stuck in the “laggard” quadrant (no current use and no intent to adopt) suggests that without intervention, spatial inequalities in technological advancement could deepen. This lends theoretical support to the idea that technology adoption is not just a firm-level phenomenon, but a geographically embedded process influenced by regional capabilities and networks \cite{porter1998competitive}.

Given the identified policy implications, future studies should explore the effectiveness of government initiatives or policy frameworks in promoting the adoption of AI. This could involve cross-state comparisons: for instance, do states (or countries) with AI strategies, tax incentives, digital infrastructure investments, or workforce training programs see faster adoption among firms? Rigorous evaluation of programs (e.g. AI innovation grants for SMEs or regional tech hubs) would provide evidence on what policy levers work best to spur adoption while mitigating inequality. Furthermore, international comparative research – contrasting the US approach with more centralized strategies elsewhere – can yield insights into how governance shapes the diffusion of AI at scale.

Another important direction is to study why certain industries and regions lag in the adoption of AI. Qualitative research in low-adoption sectors like construction or in persistently lagging states could uncover barriers such as skill shortages, regulatory hurdles, cultural resistance, or lack of relevant use cases. In contrast, examining success stories in those domains could highlight enabling factors or best practices. Comparison of leaders and laggards can reveal the role of local policies, education systems, or industry mix in facilitating AI adoption. Such findings can directly inform targeted interventions to broaden AI diffusion.

Although my study focused on adoption rates, an important next step is to examine the results of AI adoption. Future research could link adoption data with firm performance metrics (productivity, profitability, innovation output) to assess whether and how AI use translates into measurable gains. Such an analysis might involve case studies or panel data to address causality – for example, using instrumental variables or experiments to determine if AI adoption causes improvements in efficiency or if more productive firms simply adopt AI earlier, a distinct possibility given that larger and geographically diversified firms are adopting AI faster. This line of inquiry would provide evidence on the often-hypothesized productivity boost from AI and identify the conditions under which it materializes. These productivity boosts for smaller businesses are all important if AI is to ever realize its promise as a general-purpose technology.

\section{Limitations}

Although this study offers novel insights into trends in AI adoption, it also has important limitations that qualify the findings. First, the analysis is based on self-reported survey data (the Census Bureau’s BTOS), which may be subject to response biases and interpretation issues. Firms may under-report or over-report AI usage depending on their understanding of what constitutes 'AI'. The binary yes/no questions on AI use and intent do not capture the intensity or specific nature of AI adoption – for instance, occasional experimental use versus core business integration are treated the same in the data. Future research should refine measurement, possibly distinguishing between levels of AI investment or types of AI applications. 

Second, the time frame of the data is relatively short, and the trends are in the early stages. The BTOS-based analysis covers a period from late 2023 through early 2025, during which AI adoption, while increasing, remains nascent. As a result, our 'snapshot' of adoption may not generalize to longer-term patterns; inflection points or acceleration in adoption could emerge beyond the observed window. This exploratory study cannot determine whether the current linear growth will persist or whether the effects of the network will eventually spur exponential uptake. Similarly, we do not establish causal links between AI adoption and outcomes such as productivity or firm performance. The study is descriptive by design: it identifies where and among whom AI is being adopted, but not the consequences of adoption. Unobserved factors (for example, management quality, business process redesign, and complementary IT infrastructure) could influence both a firm's adoption decision and its performance outcomes. 

\bibliographystyle{unsrtnat}
\bibliography{references}  
\end{document}